\documentclass[11pt,nofootinbib,preprint]{revtex4}
\usepackage{mathrsfs}
\usepackage{graphicx}
\usepackage{amsmath}
\usepackage{amsfonts}
\usepackage{amssymb}
\usepackage{color}

\usepackage{epsfig}
\usepackage{CJK}
\usepackage{graphicx}
\usepackage{epsfig}
\usepackage{eepic}
\usepackage{bbm}
\usepackage{dcolumn}
\usepackage{bm}
\usepackage{ulem}
\usepackage{slashbox}
\usepackage{multirow}
\usepackage{slashed}

\newcommand{\omits}[1]{}

\def\bc{\begin{center}}
\def\nno{\nonumber}
\def\ec{\end{center}}
\def\be{\begin{eqnarray}}
\def\ee{\end{eqnarray}}

\definecolor{dyellow}{rgb}{1.,0.8,.0}
\definecolor{myblue}{rgb}{.1,.1,.7}
\definecolor{dcyan}{rgb}{.0,.6,.6}
\definecolor{cyan}{rgb}{0.4,1.0,1.0}
\definecolor{dmagenta}{rgb}{0.6,0.0,0.6}
\definecolor{brown}{rgb}{0.6,0.2,0.}
\definecolor{darkblue}{rgb}{.0,.0,0.5}
\definecolor{darkred}{rgb}{0.75,0.0,0.0}
\definecolor{orange}{rgb}{1.,.6,.0}
\definecolor{dorange}{rgb}{0.8,.4,.0}
\definecolor{green}{rgb}{0.0,1.0,0.0}
\definecolor{darkgreen}{rgb}{0.0,0.6,0.0}
\definecolor{purple}{rgb}{.4,.0,.4}
\definecolor{lightgrey}{rgb}{0.7, 0.7, 0.7}
\definecolor{grey}{rgb}{0.4, 0.4, 0.4}

\newcommand{\nc}{\newcommand}
\nc{\rnc}{\renewcommand} \nc{\ket}[1]{\left | \, #1 \right \rangle}
\nc{\bra}[1]{\left \langle #1 \, \right |}
\nc{\ua}{\uparrow} \nc{\da}{\downarrow}

\nc{\braket}[2]{\langle\, #1\,|\,#2\,\rangle}
\nc{\half}{\frac{1}{2}}

\nc{\prj}{\mathcal{P}} \nc{\hilb}{\mathcal{H}}
\nc{\pth}{\mathcal{C}} \nc{\inprod}[2]{\braket{#1}{#2}}
\nc{\upket}{\ket{\uparrow}} \nc{\downket}{\ket{\downarrow}}
\nc{\upbra}{\bra{\uparrow}} \nc{\downbra}{\bra{\downarrow}}

\begin{document}


\title{Note on surface growth approach for bulk reconstruction}

\author{Chao Yu$^{1}$} \email{yuch8@mail2.sysu.edu.cn}
\author{Fang-Zhong Chen$^{1}$} \email{chenfzh7@mail2.sysu.edu.cn}
\author{Yi-Yu Lin$^1$} \email{linyy27@mail2.sysu.edu.cn}
\author{Jia-Rui Sun$^{1}$} \email{sunjiarui@mail.sysu.edu.cn}
\author{Yuan Sun$^{2}$} \email{sunyuan@jlu.edu.cn}

\affiliation{${}^1$School of Physics and Astronomy, Sun Yat-Sen University, Guangzhou 510275, China}
\affiliation{${}^2$Center for Theoretical Physics and College of Physics, Jilin University,
Changchun 130012, China}



\begin{abstract}
In a recent paper, a novel surface growth approach for reconstructing bulk geometry and matter fields was proposed, it was shown that this picture can be explicitly realized by the one-shot entanglement distillation tensor network and the surface/state correspondence. In the present paper, we give direct analysis for the growth of the bulk minimal surfaces in asymptotically AdS$_3$ spacetime and show that bulk geometry can be efficiently reproduced in this way, which provides further support for the surface growth approach in entanglement wedge reconstruction.
\end{abstract}

\pacs{04.62.+v, 04.70.Dy, 12.20.-m}

\maketitle

\section{Introduction}
The anti-de Sitter/conformal field theory (AdS/CFT) correspondence has established a bridge between the boundary CFT and the gravity in the bulk asymptotically AdS spacetime~\cite{Maldacena:1997re,Gubser:1998bc,Witten:1998qj}. The correspondence also indicates an emergent picture of gravity, namely, the geometry and gravitational dynamics of bulk spacetime should in principle be constructed from information of the boundary CFT, which is called the bulk reconstruction~\cite{hamilton0606,kabat1703,jafferis1512,faulkner1704,bao1904,roy1801,faulkner1806}. In the reconstruction of bulk gravitational theory, the notion of holographic entanglement entropy plays a key role~\cite{Ryu:2006bv,Ryu:2006ef,Hubeny:2007xt}, which states that the entanglement entropy of a boundary subregion $A$ is a quarter of the area of a co-dimensional-2 minimal surface ${{\gamma _A}}$ growing into the bulk from the boundary of $A$ (to leading order in the gravitational coupling constant $G$), i.e.,
\be\label{rt}{S_A} = \frac{{\rm Area}(\gamma _A)}{4 G}.
\ee
It was shown that field theory information contained in the subregion $A$ can determine the information in spatial region bounded by $A$ and the bulk extremal surface, which is called the entanglement wedge~\cite{jafferis1512,Harlow:2016,cotler1704}. Subsequently, much progresses have been made along this direction, such as reconstruction of bulk operators from the boundary CFT operators in subregions~\cite{hamilton0606,kabat1703,jafferis1512,faulkner1704,bao1904,roy1801,faulkner1806}, generating the bulk AdS geometry from entanglement renormalization of the tensor networks~\cite{Swingle:2009bg,Swingle:2012wq,Bao:2018pvs,vidal:1812} and furthermore, investigating the emergence of gravitational dynamics from the geometry generated from the tensor networks~\cite{sun1912}. Obviously, according to RT formula (\ref{rt}), a boundary subregion $A$ can at least detect the nonlocal information about the global configuration of the extremal surface ${{\gamma _A}}$ by reading out its classical area. However, how can the information in $A$ detect (or reconstruct) the information in the region inside the RT surface ${{\gamma _A}}$ (i.e., the entanglement wedge of $A$) is not apparent and direct.

\begin{figure}[htbp]
	\begin{center}
		\includegraphics[height=7cm,clip]{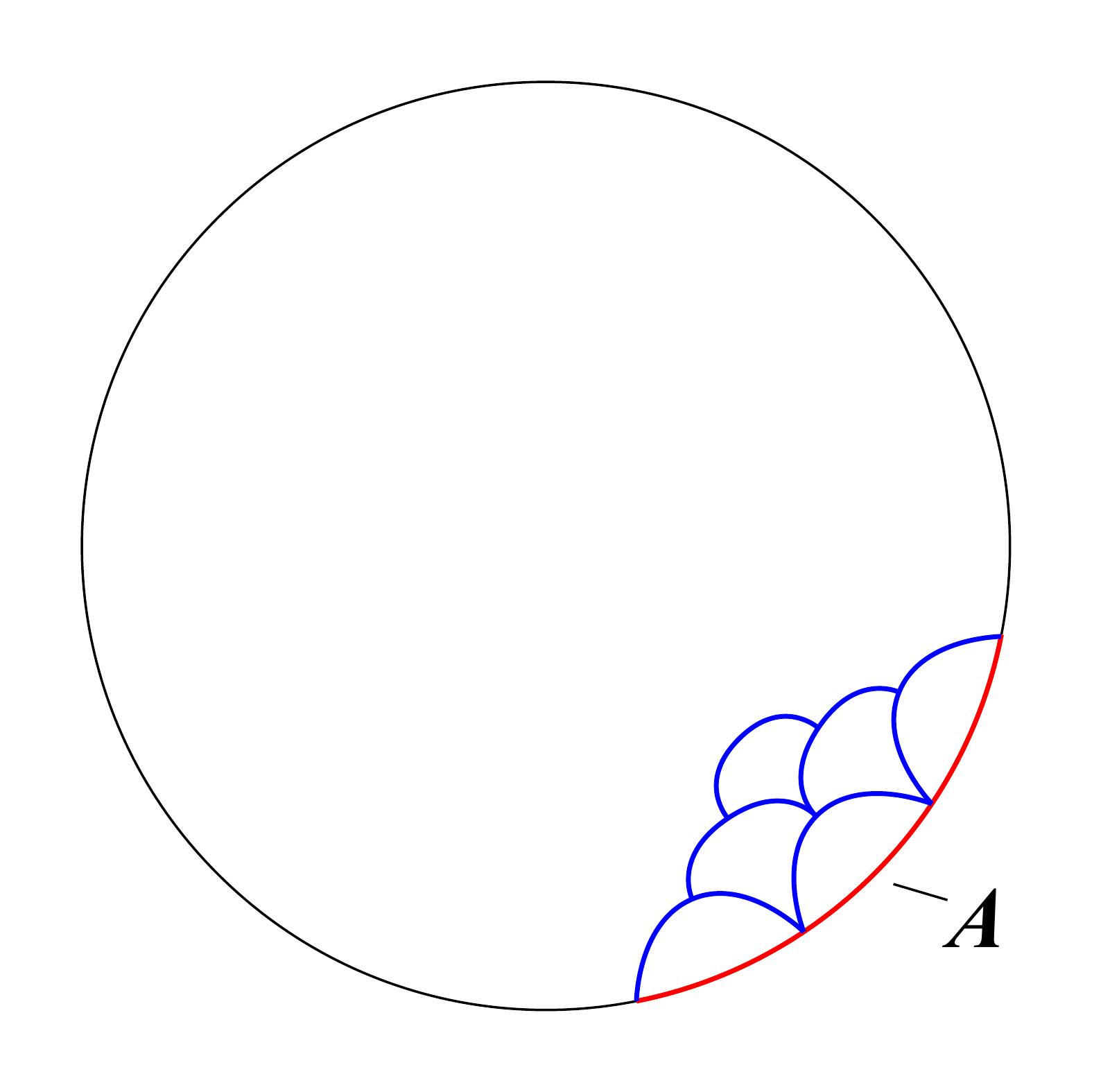}\caption{An illustration for the surface growth scheme with three layers.}
		\label{fig0}
	\end{center}	
\end{figure}

In a recent paper, three of the present authors (Lin, Sun and Sun) proposed a concrete and very natural approach to reconstruct the bulk geometry in the entanglement wedge from a surface growth procedure, similar to the Hygens' principle of wave propagation. The basic ideas of the approach can be briefly described as follows~\cite{Lin:2020thc}. Firstly, considering there are many minimal surfaces ``growing'' out from a set of small boundary subregions side by side. Secondly, regarding these minimal surface as the new boundaries and further take the points on them as the anchor points for the new minimal surfaces to grow into the deeper bulk regions. Then repeating the procedure layer by layer such that the new minimal surfaces can probe arbitrary region in the entanglement wedge. Consequently, the information of these bulk regions can be detected and reconstructed from the initial boundary regions in this way, see FIG.~\ref{fig0}. A closely related case is the horizon entropy of the brane world black hole, in which its horizon will extend into the bulk spacetime and it is just a minimal surface when the black hole is stationary, and then the black hole Bekenstein-Hawking entropy can be naturally interpreted as the holographic entanglement entropy between regions in and outside the black hole horzion~\cite{Emparan:2006ni}. It was shown that the above picture can be explicitly realized by generalizing the one-shot entanglement distillation (OSED) method and the surface/state correspondence~\cite{Bao:2018pvs,miyaji:1503}, and the well-known MERA-like tensor network in \cite{vidal:2007,Vidal:2008,vidal:1812} can be exactly identified with specific surface growth configurations~\cite{Lin:2020thc}, thus provides a concrete and intuitive way for the entanglement wedge reconstruction~\cite{jafferis1512,Harlow:2016,cotler1704}.

Since the surface growth picture is a general geometric description which is independent of the tensor network, therefore, in the present paper, we will further analyze the surface growth approach in asymptotically AdS$_3$ spacetime from calculating the minimal surfaces directly. We will study both the homogeneous and inhomogeneous subsystem cases and will show that the spatial region inside the entanglement wedge can be efficiently reconstructed, like bubble growing, which presents a clear process of boundary to bulk propagation.

\section{A tensor network viewpoint of the surface growth scheme}
In this section, we briefly review our previous work~\cite{Lin:2020thc}, which provides an interesting tensor network (more explicitly, the so-called OSED tensor network) perspective on the surface growth scheme.

Tensor network was originally used as a numerical simulation tool to effectively represent quantum many-body states in condensed matter physics. It is characterized by relating a bulk graph of some geometric structure with the entanglement structure of a quantum system. Later on, some specific types of tensor network models were tentatively used to describe the mechanism of the holographic duality~\cite{Bao:2018pvs,Lin:2020thc,Swingle:2009bg,Swingle:2012wq,Milsted:2018san,Pastawski:2015qua,Hayden:2016cfa,Qi:2013caa}.

\begin{figure}[htbp]     \begin{center}
		\includegraphics[height=8cm,clip]{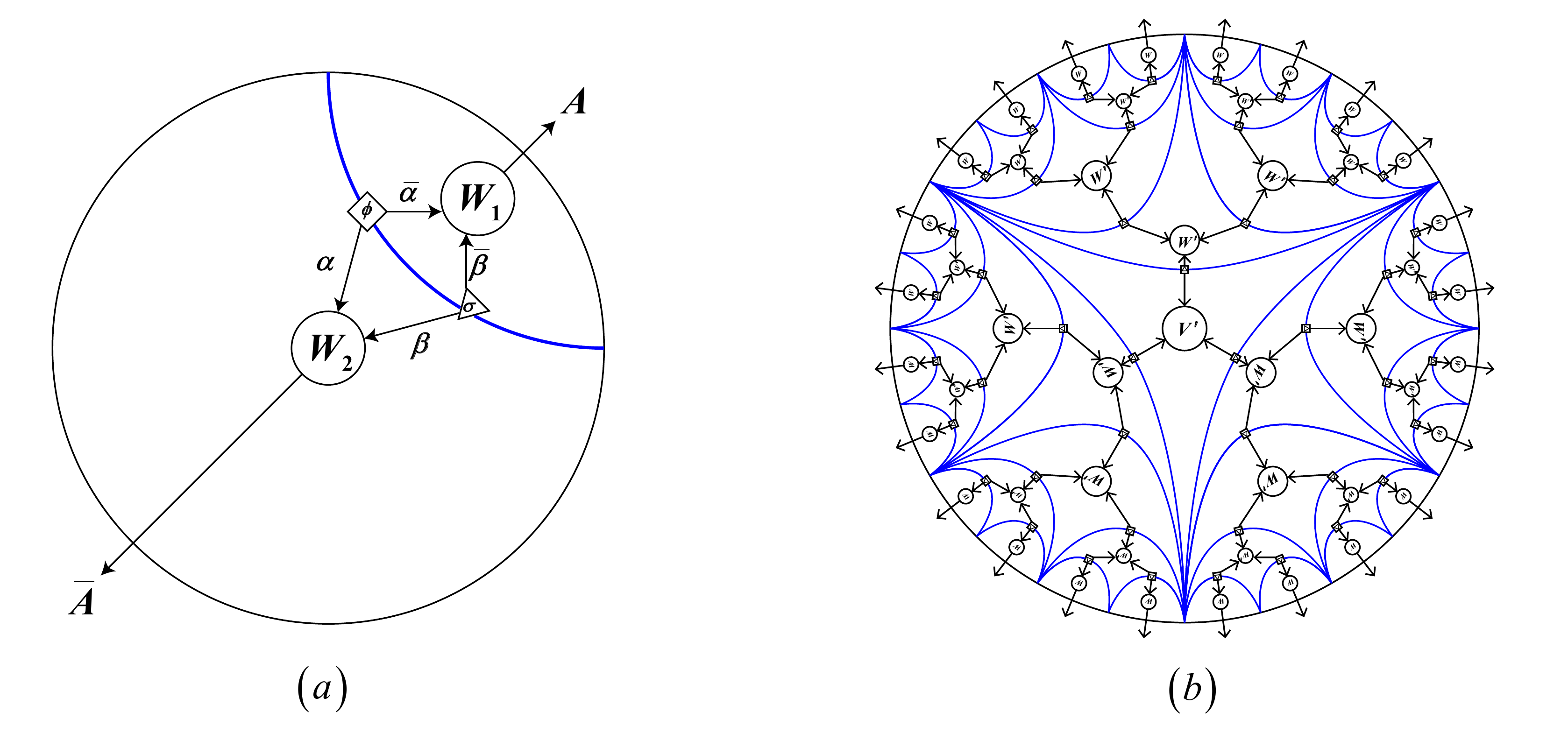}
		\caption{(a) The simplest OSED tensor network for a bipartite holographic state. (b) Another example: a special OSED tensor network with spherical symmetry and fractal feature. It can be identified with the well-known MERA-like tensor network.}
		\label{fig re}
	\end{center}	
\end{figure}

The OSED (one-shot entanglement distillation) tensor network is a tensor network model constructed on the basis of the OSED procedure proposed in~\cite{Bao:2018pvs}. Let us first briefly describe the concept of OSED. It was shown that due to the holographic limits, one can always construct the so-called ``smoothed states'' $\left| {{\Psi ^\varepsilon }} \right\rangle$ and $\rho _A^\varepsilon$ to approximate the given holographic CFT full, pure state $\left| \Psi  \right\rangle $ and the reduced density matrix ${\rho _A}$ for a certain subregion $A$ of the CFT respectively. Then, rearranging the eigenvalues of the smoothed state $\rho _A^\varepsilon $ in descending order and dividing them into blocks of size $\Delta  = {e^{{S_A} - O(\sqrt {{S_A}} )}}$, one can further approximate the boundary state $\left| \Psi  \right\rangle $ as the following tensor network representation (see more details in~\cite{Bao:2018pvs,Lin:2020thc}):
\be\label{def0}{\Psi ^{A \bar A}} = {W_1}_{\bar{\beta} \bar{\alpha}}^{A}{W_2}_{\beta \alpha }^{\bar A}\phi ^{\alpha \bar \alpha}{\sigma ^{\beta \bar \beta}} = ({W_1} \otimes {W_2})(\left| \phi  \right\rangle \otimes \left| \sigma  \right\rangle), \ee
where a maximally entangled state $\left| \phi  \right\rangle \otimes \left| \sigma  \right\rangle$ has been distilled out, which is defined by
\be\label{def1}\begin{array}{l}
	\left| \phi  \right\rangle  = \sum\limits_{m = 0}^{{e^{S - O(\sqrt S )}}} {{{\left| {m\bar m} \right\rangle }_{\alpha \bar \alpha }}}, \\
	\left| \sigma  \right\rangle  = \sum\limits_{n = 0}^{{e^{O(\sqrt S )}}} {\sqrt {\tilde p _{n\Delta }^{{\rm avg}}} {{\left| {n\bar n} \right\rangle }_{\beta \bar \beta}}},
\end{array}\ee
where $S$ denotes the entanglement entropy of $A$, and ${{\tilde p _{n\Delta }}^{\rm avg}}$ is the average eigenvalue of each block. Clearly, the logarithm of the Hilbert space dimension of $\left| \phi  \right\rangle $ matches the entanglement entropy of $A$, and $\left| \sigma  \right\rangle $ act as the quantum fluctuation. The tensors ${W_1}$ and ${W_2}$ are isometries, which map the auxiliary states represented by the bonds of ${\phi ^{\alpha\bar \alpha}}{\sigma ^{\beta \bar \beta}}$ into the eigenstates of the reduced density matrices for $A$ and $\bar A$ respectively. Eq.(\ref{def0}) is the OSED for a holographic state, see FIG.~\ref{fig re}(a).

Then, utilizing a series of nonintersecting RT surfaces according to a certain appropriate order to discretize the bulk spacetime into cells in the same way, ref.~\cite{Bao:2018pvs} showed that by iterating the OSED procedure on a holographic boundary state, one can assign an isometry tensor for each cell to implement the map between the states associated with its minimal surface boundaries, and finally construct a whole OSED tensor network, which can reproduce the correct boundary state with high fidelity, and have a bulk geometry that matches with the bulk AdS spacetime perfectly. Interestingly, in~\cite{Lin:2020thc}, we construct a special OSED tensor network with spherical symmetry and fractal feature in this way, see FIG.~\ref{fig re}(b), and we demonstrate that it can be identified with the well-known MERA-like tensor network.

Moreover, in~\cite{Lin:2020thc}, we further investigated the physical meaning of the OSED tensor network. More specifically, a mixed state $\{ (\left| {\bar m} \right\rangle ,{p_m} = \frac{1}{{{e^S}}})\}$ with equal probabilities to each extremal surface in the bulk is assigned, since in the framework of surface/state correspondence the density matrix corresponding to an extremal surface is a direct product of density matrices at each point. Therefore, the isometry tensor in the OSED tensor network actually plays the role of implementing the entanglement distillation procedure on the state of the union of the minimal surfaces in the previous layer and then to map it into the state on the next minimal surfaces. This understanding leads to the generalization of the OSED tensor network into the cases involving more general bulk minimal surfaces whose anchor points are located on the previous bulk minimal surface. Therefore, the generalized OSED tensor network provides a natural interpretation for our surface growth scheme, namely, the emergence of bulk spacetime geometry can be interpreted as layers of minimal surfaces continue to grow into the deeper bulk regions from the initial boundary regions.

\section{The growth of minimal surface in pure AdS$_3$}
Now let us move on to the direct growth of bulk minimal surface in pure AdS$_3$ spacetime in global coordinate
\be\label{globalads3}
ds^2=d\rho^2+L^2\left(-\cosh^2\frac{\rho}{L}dt^2+\sinh^2\frac{\rho}{L}d\phi^2\right),
\ee
where $L$ is the curvature radius of the AdS spacetime. The bulk static co-dimensional-2 surface (which is a curve in AdS$_3$ case) can be expressed as $\rho=\rho(\phi)$. Then the length of the curve is
\be\label{length}
\gamma &=&\int\sqrt{\left(\rho'^2+L^2\sinh^2\frac{\rho}{L}\right)}d\phi\nno\\
&\equiv& \int\mathcal{L}d\phi,
\ee
where $\rho'=\frac{d\rho}{d\phi}$. The bulk minimal surface satisfies the Euler-Lagrange equation
\be
\frac{\partial \mathcal{L}}{\partial\rho}-\frac{d}{d\phi}\frac{\partial\mathcal{L}}{\partial\rho'}=0,
\ee
which gives
\be\label{miniads3}
\tilde{\rho}''\sinh\tilde{\rho}-2\tilde{\rho}'^2\cosh\tilde{\rho} -\sinh^2\tilde{\rho} \cosh\tilde{\rho} =0,
\ee
where $\tilde{\rho}\equiv\rho/L$. Eq.(\ref{miniads3}) can be solved as
\be\label{soluads}
\phi &=&\pm\arctan\left(\frac{\sinh^2\tilde{\rho}}{\sinh\tilde{\rho}_*}
+\cosh\tilde{\rho}\sqrt{\frac{\sinh^2\tilde{\rho}}{\sinh^2\tilde{\rho}_*}-1}\right)\nno\\
&&\mp \arctan\left(\sinh\tilde{\rho}_*\right)+\phi_0,
\ee
in which $\rho_*$ is the turning point of the bulk geodesic and $\phi_0=\phi(\rho_*)$. Note that the value of $\rho_*$ will change for different bulk geodesics. Substituting eq.(\ref{soluads}) into eq.(\ref{length}), then the length of the geodesic is
\be\label{length2}
\gamma=L\ln\frac{\left(\cosh\tilde{\rho}_1+\sqrt{\sinh^2\tilde{\rho}_1-\sinh^2\tilde{\rho}_*}
\right)\left(\cosh\tilde{\rho}_2+\sqrt{\sinh^2\tilde{\rho}_2-\sinh^2\tilde{\rho}_*}
\right)}{\cosh^2\tilde{\rho}_*},
\ee
where we require that $\tilde{\rho}_*<\tilde{\rho}_1<\tilde{\rho}_2$.

\begin{figure}[htbp]     \begin{center}
		\includegraphics[height=9cm,clip]{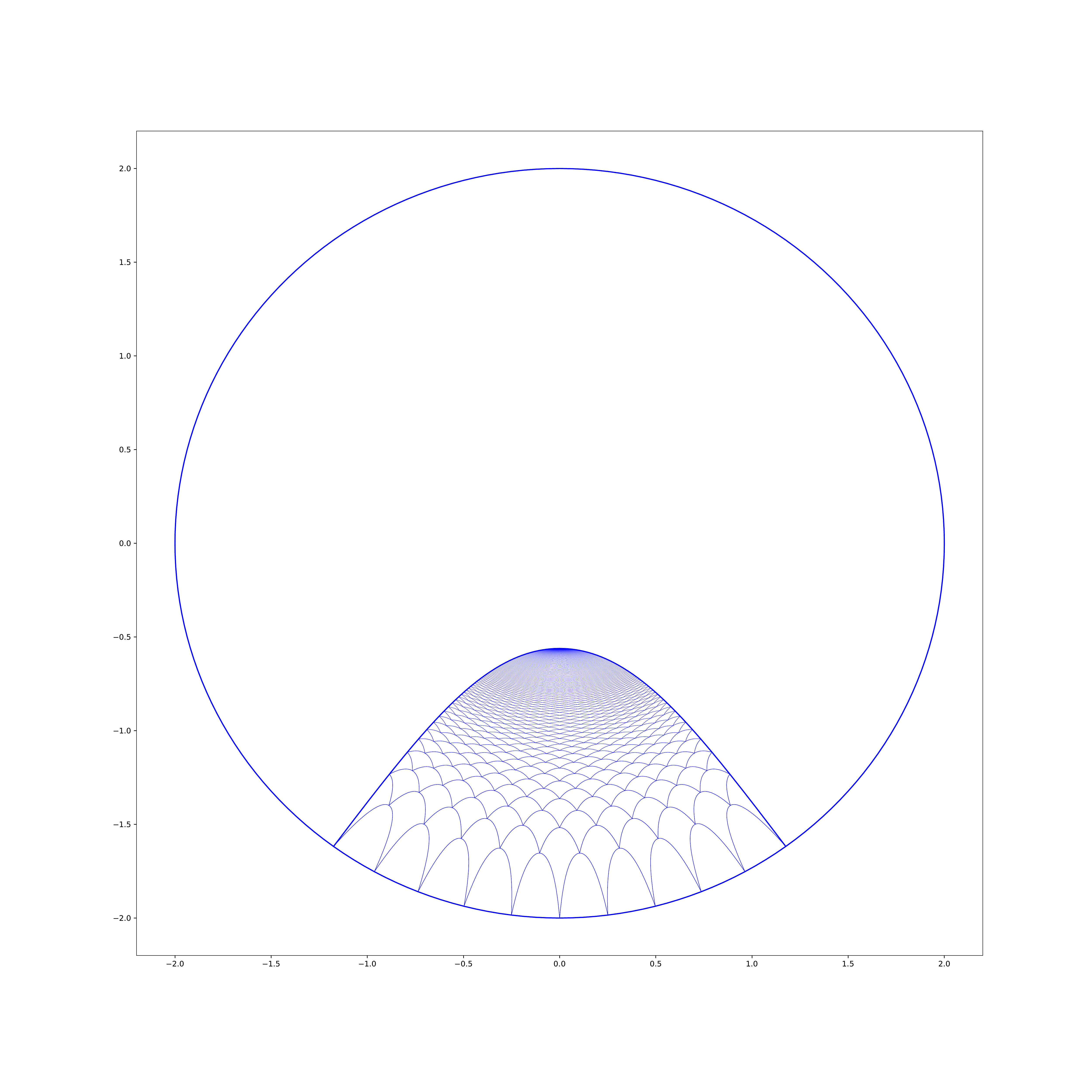}\caption{Surface growth for initial ten subsystems with equal spatial size, where each subsystem expands an angle $\phi=\pi/25$, and the conformal boundary is chosen as $\rho_0=2$ and $L=0.5$, the growing steps are 300.}
		\label{ads3sym}
	\end{center}	
\end{figure}

Now we can draw the surface growth picture as in FIG.~\ref{ads3sym}. At the conformal boundary, we firstly chose ten subsystems with equal spatial size, each of them expands an angle $\phi=\pi/25$, after the growth of the first layers of the geodesics, we chose the boundaries of the second layers of geodesics to be at the centers of the first ones, then the bulk space within the entanglement wedge can be filled after finite steps. An interesting point is that, the outer bulk geodesic which corresponds to the whole ten subregions also should be treated as the spatial boundary when considering the surface growth. Otherwise, the region growing from the ten subregions will be smaller than the outer entanglement wedge.

\begin{figure}[htbp]     \begin{center}
		\includegraphics[height=9cm,clip]{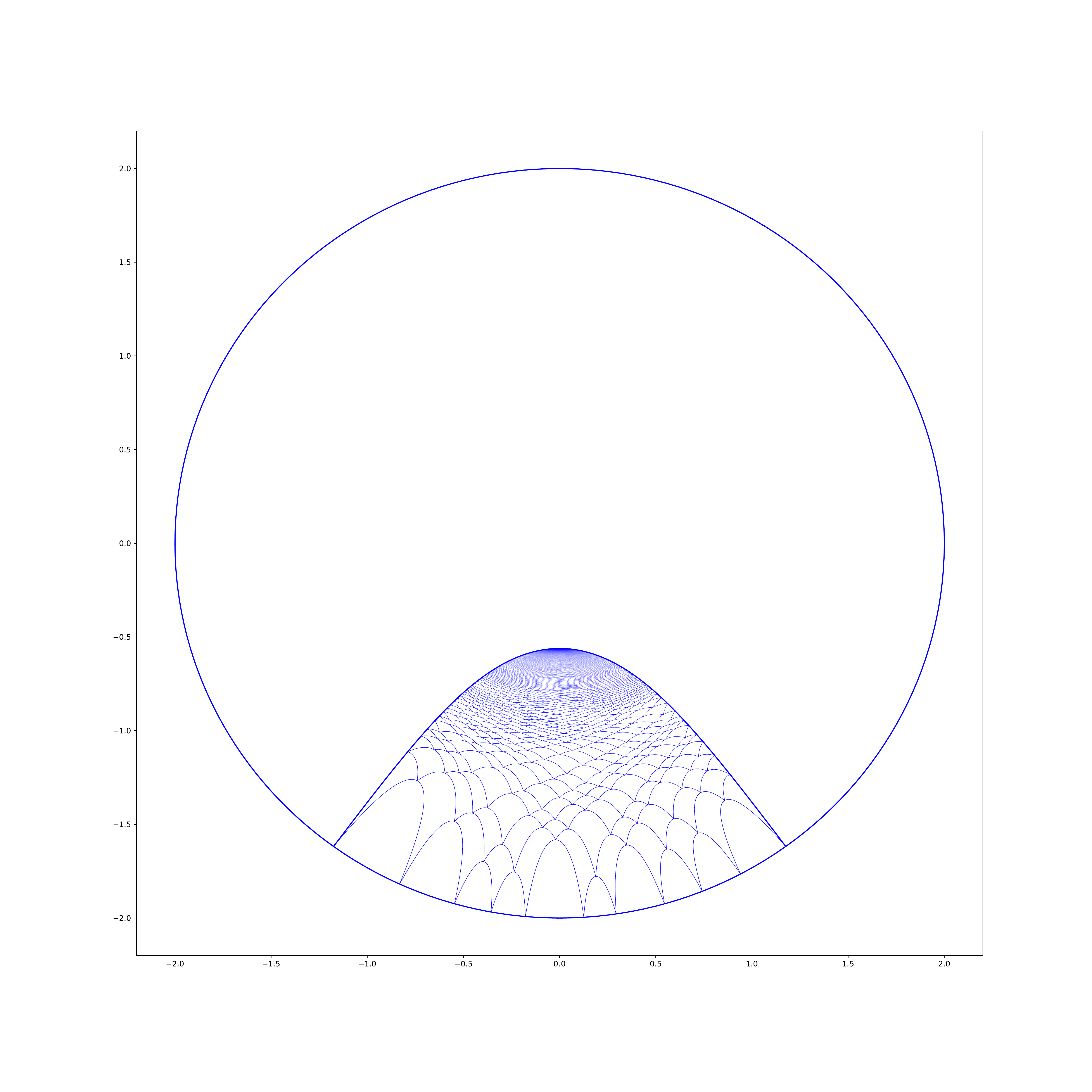}\caption{Surface growth for initial ten subsystems with different spatial sizes, and the conformal boundary is chosen as $\rho_0=2$ and $L=0.5$, the growing steps are 300.}
		\label{ads3asym}
	\end{center}	
\end{figure}

We can also consider the surface growth for ten subregions with arbitrary sizes on the boundary, see FIG.~\ref{ads3asym}. Since the subregions are of different sizes, from the second layer on, some minimal surfaces become asymmetric, thus when considering the new geodesics connecting two adjacent geodesics, we chose the the one with maximal value. Like the symmetric case in FIG.~\ref{ads3sym}, the outer boundary (the minimal surface corresponding to the whole subregions) is necessary for the growing minimal surfaces to fill the entanglement wedge.

\begin{figure}[htbp]     \begin{center}
		\includegraphics[height=9cm,clip]{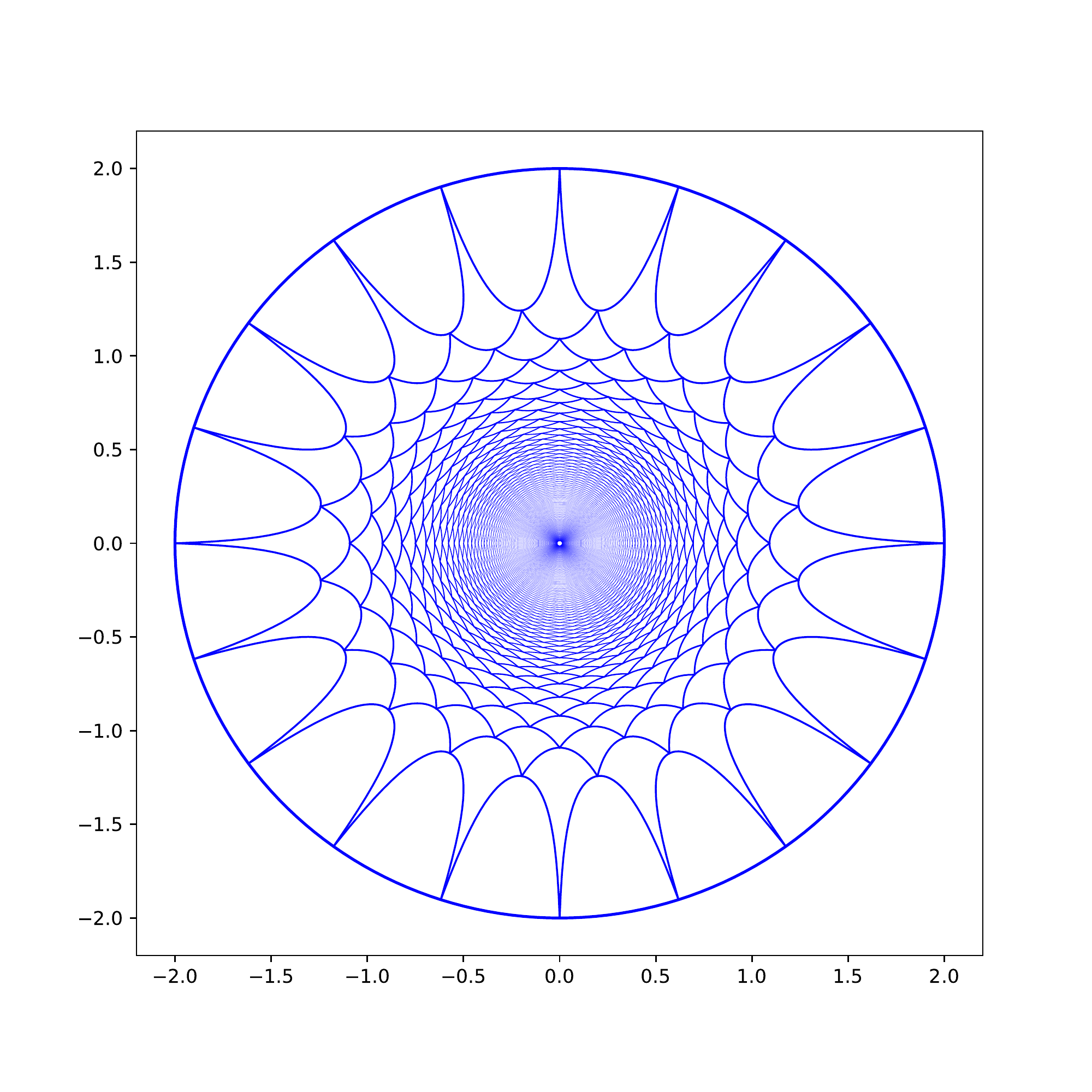}\caption{Surface growth from the whole boundary when they are divided into twenty subsystems with equal spatial size, where each subsystem expands an angle $\phi=\pi/10$, and the conformal boundary is chosen as $\rho_0=2$ and $L=0.5$, the growing steps are also 300.}
		\label{adsfull}
	\end{center}	
\end{figure}

Another interesting case is to construct the full AdS space from the surface growth. To achieve this, we can divide the whole boundary into many subregions, without loss of generality, we choose twenty subregions with identical size as in FIG.~\ref{adsfull}, and choose the boundaries of the next layers of minimal surfaces to be the centers of the previous layers. Then after finite steps of minimal surface growing, the full AdS space will be filled or constructed.
%

\section{The growth of minimal surface in nonrotating BTZ black hole}
In this section we will study the bulk reconstruction from surface growth in nonrotating BTZ black hole. The metric of the nonrotating BTZ black hole is~\cite{Banados:1992gq,Banados:1992wn}
\be\label{btz}
ds^2=-\frac{r^2}{L^2}f(r)dt^2+\frac{L^2}{r^2 f(r)}dr^2+r^2 d\phi^2,
\ee
where $f(r)=1-\frac{M}{r^2}$ and its horizon is located at $r_h=\sqrt{M}$. The bulk static curve is $r=r(\phi)$, which has the length as
\be\label{lengthbtz}
\gamma=\int d\phi\sqrt{g(r)\dot{r}^2+r^2},
\ee
where $g(r)=\frac{L^2}{r^2 f(r)}$ and $\dot{r}=\frac{dr}{d\phi}$. The bulk geodesic is determined by the Euler-Lagrange equation, which gives
\be\label{minibtz}
2r^2+4gu^2-rg'u^2-2gruu'=0,
\ee
with $u=u(\phi)\equiv \dot{r}$ and $g'=\frac{dg}{dr}$. Eq.(\ref{minibtz}) can be solved as
\be
u(\phi)\equiv \frac{dr}{d\phi}=\pm\sqrt{\frac{r^4-r^2r_*^2}{r_*^2g(r)}},
\ee
where $r_*$ indicates the turning point of the bulk geodesic. Subsequently, we can obtain the solution
\be\label{minibtzsol}
\phi-\phi_0=-\frac{L}{r_h}\ln\left(\sqrt{1-\frac{r_h^2}{r^2}}-\sqrt{\frac{r_h^2}{r_*^2}-\frac{r_h^2}{r^2}}\right)
+\frac{L}{r_h}\ln\sqrt{1-\frac{r_h^2}{r_*^2}},
\ee
where $\phi_0=\phi(r_*)$ denotes the center of the bulk geodesic. Consequently, the length for the bulk geodesic is
\be\label{lengthbtz2}
\gamma=L\cosh^{-1}\left(\frac{2r_1^2-(r_h^2+r_*^2)}{r_*^2-r_h^2}\right)+L\cosh^{-1}\left(\frac{2r_2^2-(r_h^2+r_*^2)}{r_*^2-r_h^2}\right),
\ee
in which we require that $r_*<r_1<r_2$ and $\phi(r_1)<\phi_0<\phi(r_2)$.

\begin{figure}[htbp]     \begin{center}
		\includegraphics[height=9cm,clip]{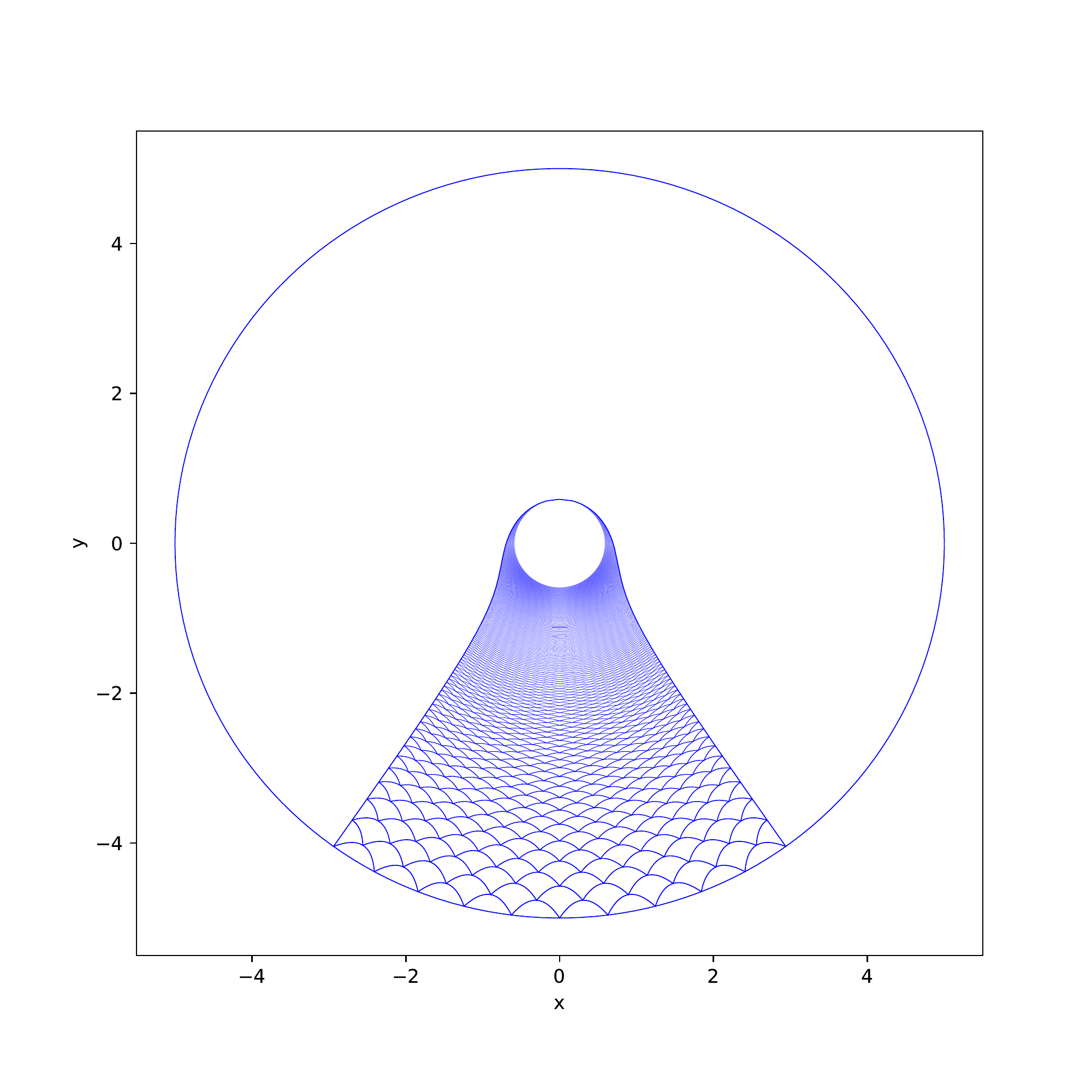}\caption{Surface growth from ten boundary subregions with equal size, where each subsystem expands an angle $\phi=\pi/25$, and the conformal boundary is chosen as $r_0=5$ and $L=1$, the growing steps are 360.}
		\label{btz10}
	\end{center}	
\end{figure}

Let us consider the surface growth from ten boundary subregions with equal size and each subregion expands an angle of $\pi/25$, and also choose the boundaries of the next layers of minimal surfaces to be the centers of the previous layers.. As can be seen from FIG.~\ref{btz10} that, due to the existence of black hole horizon, the growing minimal curves will surround the horizon after finite steps of growing. Of course, if the total size of the whole subregions is not large enough, the growing geodesics will not reach the black hole horizon.

\begin{figure}[htbp]     \begin{center}
		\includegraphics[height=9cm,clip]{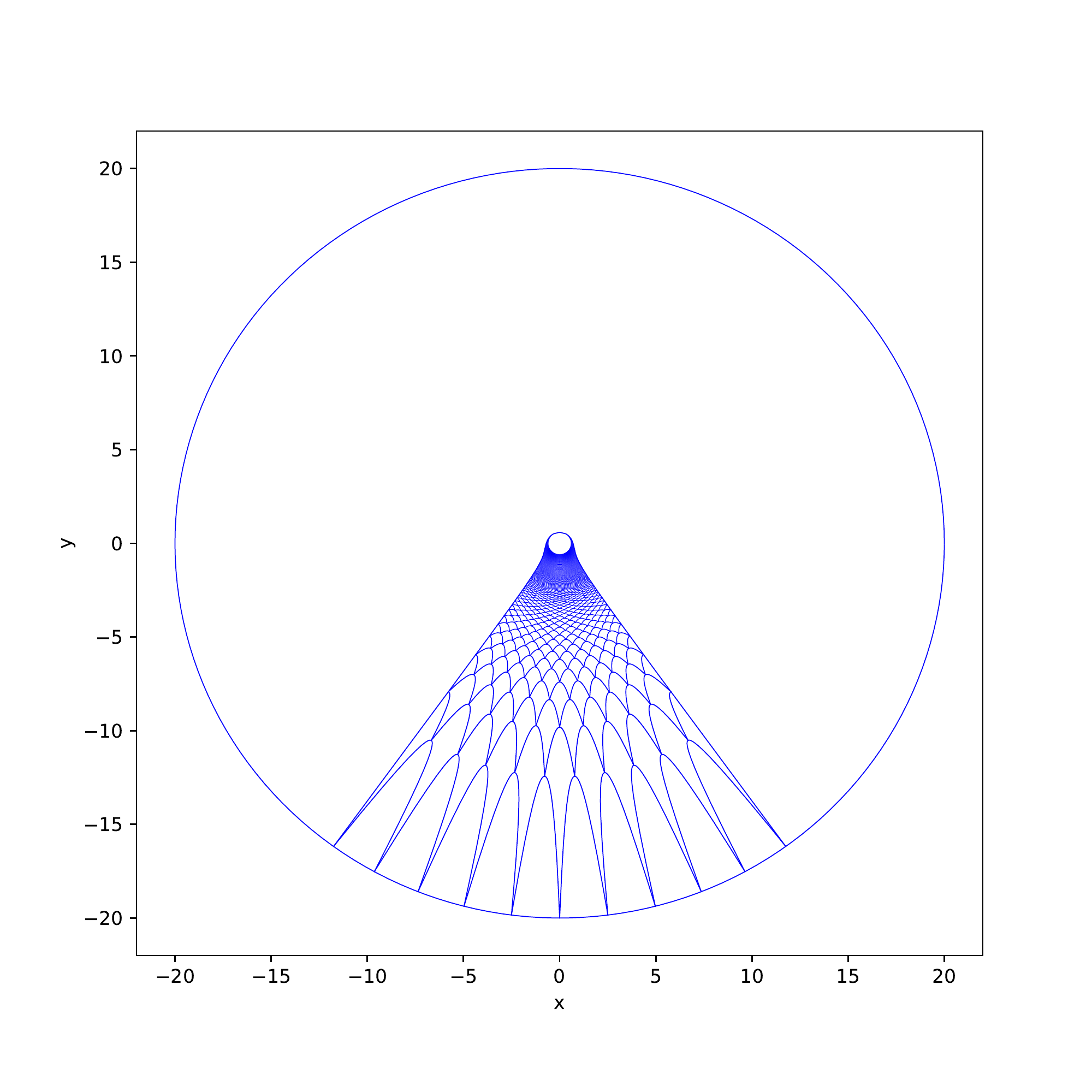}\caption{Surface growth from ten boundary subregions with equal size, where each subsystem expands an angle $\phi=\pi/25$, and the cutoff surface is chosen as $r_0=r_c=20$ and $L=1$, the growing steps are also 360.}
		\label{btzr20sym}
	\end{center}	
\end{figure}

Note that the asymptotic boundary of the AdS spacetime is located at $r\rightarrow\infty$. However, this will not affect the bulk reconstruction from the surface growth. Since the surface growth process actually corresponds to the holographic renormalization group flow, or the MERA of the corresponding tensor network~\cite{Lin:2020thc}. Therefore, the bulk minimal surfaces can grow from any finite cutoff surfaces $r=r_c$. For example, we can choose the cutoff surface to be located at $r_0=r_c=20$, and also consider ten homogeneous subregions, see FIG.~\ref{btzr20sym}. By comparing FIG.~\ref{btzr20sym} with FIG.~\ref{btz10} we can see that after very few steps of surface growing, the minimal surfaces in FIG.~\ref{btzr20sym} will reach the region $r_c=5$.

\begin{figure}[htbp]     \begin{center}
		\includegraphics[height=9cm,clip]{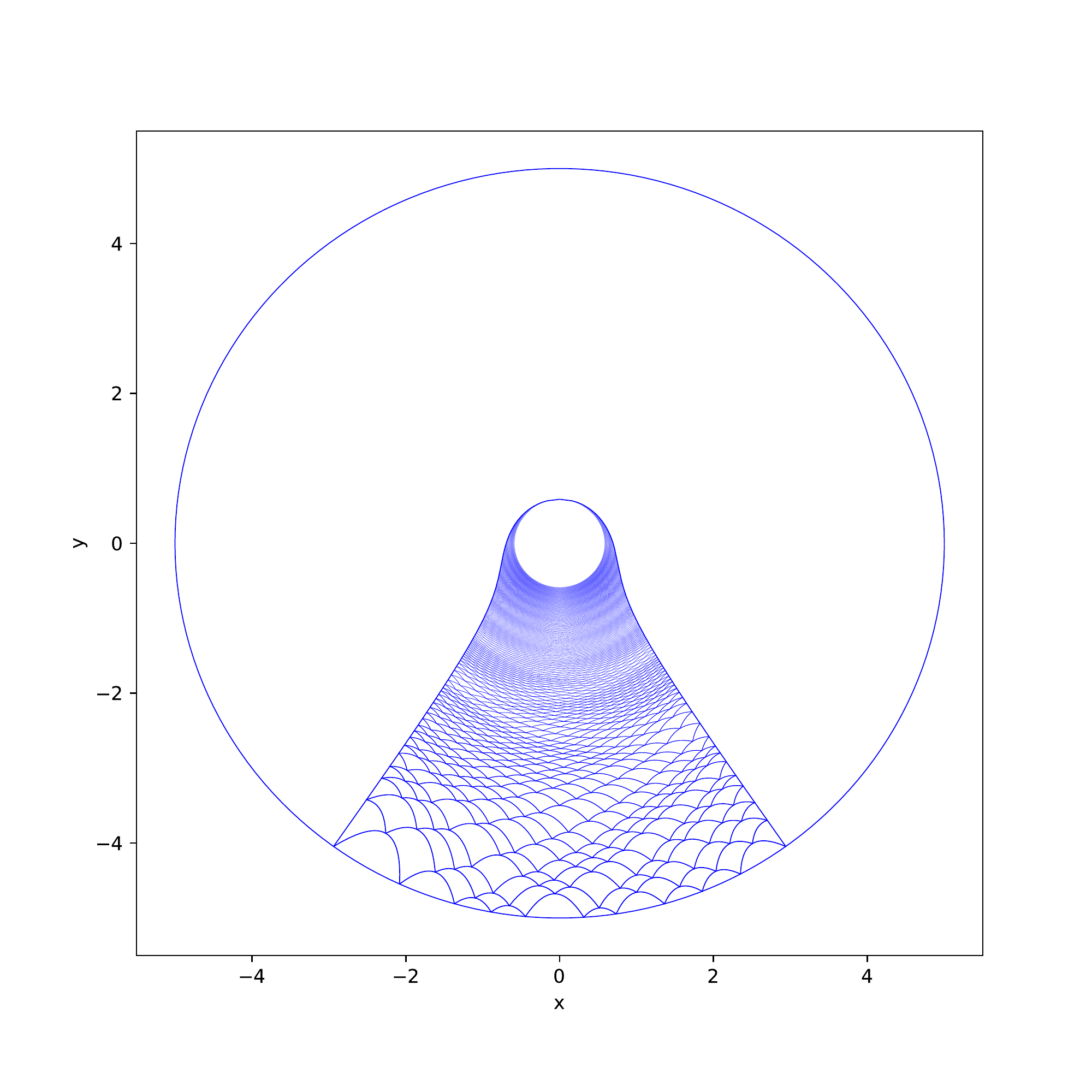}\caption{Surface growth from ten inhomogeneous boundary subsystems, where whole subsystems expand an angle $\phi=2\pi/5$, and the cutoff surface is chosen as $r_0=r_c=5$ and $L=1$, the growing steps are 358.}
		\label{btzr10asym}
	\end{center}	
\end{figure}

As a final example, we will consider the surface growth from ten inhomogeneous subsystems at cutoff surface $r_c=5$, see FIG.~\ref{btzr10asym}. Again, from the second layer on, the boundary points of some of the geodesics cannot end on the middle of the previous layers, instead, we need to choose the maximal one connecting the adjacent two geodesics in the previous layer. In addition, it can be seem from FIG.~\ref{btzr10asym} that as the geodesics growing deeper into the BTZ black hole, the surface growth configuration becomes more homogeneous.

%

\section{Conclusions and discussions}\label{sect:conclusion}
In this paper, we further investigated the surface growth approach for bulk reconstruction from analyzing the growth of bulk minimal surfaces (geodesics) in asymptotically AdS$_3$ spacetime directly. We showed in various cases that the spatial region in the outer entanglement wedge can be constructed from the growth of the bulk minimal surfaces layer by layer, which gives further support for the surface growth approach and presents a clear picture of the boundary to bulk propagation and subregion to subregion duality in the gauge/gravity duality. In addition, in the surface growth scheme, each previous layer of growing minimal surfaces act as the new boundary for the next layer, this is in accord with the holographic Wilsonian renormalization group flow, in which the effective field theory can be located at any radial cutoff surface from UV to IR regions~\cite{Faulkner:2010jy,Heemskerk:2010hk}. Moreover, the surface growth process corresponds to the extended OSED tensor network, and it was also shown to be closely related to the entanglement of purification and bit threads description for bulk reconstruction~\cite{Lin:2020yzf,Lin:2021hqs}. The surface growth approach provides a concrete and intuitive realization of the subregion duality and an efficient way for the reconstruction of bulk geometry and matter fields, there are many interesting problems to explore for the surface growth approach such as studying its relation with the $T\bar{T}$ deformation in the dual CFT~\cite{McGough:2016lol,Guica:2019nzm} and including the perturbative corrections during the surface growth~\cite{He:2014lfa}.

\section*{Acknowledgement}
We would like to thank L.-Y. Hung for helpful discussions. J.R.S. was supported by the National Natural Science Foundation of China (No.~11675272). This work was also supported by the National Natural Science Foundation of China (No. 12105113) and China Postdoctoral Science Foundation (No. 2019M653137).





\end{document}